\documentclass[12pt]{article}
\usepackage{epsfig}
\usepackage{graphicx}
\tt
\begin{document}
\rm
\begin{center}
\Large\bf On the Planetary acceleration and the Rotation of the
Earth\end{center} \hspace{3cm}
\begin{center}
\textbf{Arbab I. Arbab}\footnote{E-mail: aiarbab@uofk.edu}\\
\vspace{1cm}
 Department of Physics, Faculty of Science, University of Khartoum, P.O. Box 321, Khartoum 11115, Sudan
\end{center}
\begin{center}
\author{Arbab I. Arbab}
\end{center}
\begin{abstract}
We have developed a model for the Earth rotation that gives a good
account (data) of the Earth astronomical parameters. These data
can be compared with the ones obtained using space-base
telescopes. The expansion of the universe has an impact on the
rotation of planets, and in particular, the Earth. The expansion
of the universe causes an acceleration that is exhibited by all
planets.
\end{abstract}
\section{Introduction}

It has been understood that the impact of the universe expansion
on our solar system is negligible. This is however not very true.
The consequences of the expansion on the earth - moon system is in
the measurable limit. The evolution of the earth-moon system was
understood to be mainly due to tidal evolution. We have recently
shown (Arbab, 2003) that the present acceleration of the universe
is due to the ever increasing gravity strength. Very recently, we
have found that the evolution of angular momenta and energy of the
earth-moon system can be accounted as due to cosmic expansion
(Arbab, 2005). This system is affected by the perturbation due to
other planets or the sun. The cosmic expansion may show up in
raising tides in this system. The influence of the expansion is
contained in changing the value of the gravitational constant
appearing in Kepler's third law and Newton's law of gravitation.
At any rate, the total cosmic effect is embedded in the an
effective gravitational constant ($G_{\rm eff.}$) that takes care
of any gravitational interactions with thee system. For a flat
universe if gravity strengthens, expansion has to increase, in
order to maintain a flatness condition. If gravity increases, then
its effect on rotation of the earth - moon system will show up in
its evolution. Astronomical investigations show that the preset
earth's rotation is decreasing, so that the length of the day is
increasing at a rate of 2 millisec/century. Astronomical analysis
could not account for the entire rotation of the earth. For the
time being, one can only extrapolate, which might be too
dangerous. Laser ranging experiments (LRE) show that the moon is
receding as it acquires an angular momentum due to the spin down
of the earth rotation. That is because the total angular momentum
of the earth - moon system remains constant during its evolution.
Moreover, the moon  exhibits some kind of an anomalous
acceleration, that can be measured.

We however provide here a different approach to study this system.
The data obtained are in agreement with geophysical and
palaeontological findings. We attribute this evolution to the
cosmic acceleration. However, the effect might appear in making
tides. Recent findings based on optical observations in the solar
system suggest that all planets might accelerate in their orbits.
About thirty years ago the first indications of planetary drifts
away from their predicted ephemerides appeared in the literature
and more recently (Kolesnik, 2000) reported that that planetary
drifts, determined from optical observations, may possibly be
accelerations proportional to their motions.

\section{The Model}

We have recently developed a model that accounts for the present
cosmic acceleration (Arbab, 2003). We have shown that, in the
present epoch, the gravitational constant ($G$) increases with
time . However, its exact time dependence is not well determined
form cosmology. One has to resort to other source of information.
This is found to be the past earth rotation.

It is known that the earth rotation is decreasing with time since
the earth was formed. Scientists attribute this to the tide rasing
force by the moon on earth. Accordingly, the day is lengthening at
a a rate of about 2 millisecond every century. Hence, the earth is
losing angular momentum and the moon must increase its angular
momentum, as due to the angular momentum conservation of the
earth-moon system. This fact implies that the moon must be
receding from the earth. We know that the motion of the earth
around the sun conserves the angular momentum. One can satisfy
this conservation by requiring the earth to accelerate in its
orbit around the sun. According to the scale expanding cosmos
(SEC) the present acceleration of the earth is about 2.8 arcsec
per century squared (Kolesnik  and Masreliez, 2004). Moreover, one
can attribute the deceleration of the earth rotation as due to
cosmic expansion. The variation of the length of day are normally
thought as due to the tidal dissipation raised by the moon on the
earth. Others connect this deceleration with the interactions of
the Earth core. However, in the present scenario we only know the
total contribution, which we trust to be a consequence of the
acceleration of the universe. This accelerated expansion is
counteracted by a growing gravitational force between celestial
objects. This gradual increase in gravity force is the main
consequence of the astronomical phenomena we now come to observe.
Geologists observed that the length of the day has not been
constant over the past million years. Besides, they observe a
similar change in the number of days in a year, days in a month,
distance between earth and moon. These variations can be
calculated and their corresponding values can then be confronted
with observational data.

We suggest that the cosmic expansion has an influence on the
Earth-Sun-Moon system and similar systems. For a bound system,
like the Earth-Sun, to remain in a bound state, despite the cosmic
expansion (possibly accelerating), gravity strength has to
increase to compensate for the cosmic expansion consequences. This
strengthening of gravity would manifest its self in some aspects,
like tidal acceleration, or orbital acceleration. We anticipate
the Earth-Sun distance to change with cosmic time too. This means
in the remote past the planets were at different positions from
the Sun when they were formed.

The viability of this model will depend on the future astronomical
or geological data that will emerge thereafter. The formulae we
have obtained are not extrapolation, but rather emerge originally
from a Gravitational Theory of Relativity (GTR), and therefore are
reliable. They represent empirical relations that account for the
rotation and evolution of the Earth-Sun system and similar system.
Present data can not be used to understand the full history of the
Sun-Earth-Moon system by just extrapolating them over very distant
past. Hence, the use of our data will be inventible. Our model is
so far the only model that provides a temporal evolution of the
Earth-Moon-Sun system parameters. The prediction of these formulae
are overwhelming, however. Theoretical prejudice favors that the
Earth primordial rotation is about six hours. Only our model can
give this value.

From the angular momentum $(L)$ and the Kepler's third law, one
finds
\begin{equation}\label{1}
L^3\propto G^2 T\ ,
\end{equation}
\begin{equation}\label{1}
L\propto \sqrt{Gr}\ ,
\end{equation}
and
\begin{equation}\label{1}
L\propto G v^{-1}\ ,
\end{equation}
where $T$ is the number of days in a year, $v$ is the orbital
velocity of the earth (planet) rotation in its orbit, $r$ is the
earth (planet)-sun distance, and $G$ is the gravitational
constant. If the angular momentum of the earth-Sun system is
constant, then on find that
\begin{equation}\label{1}
T\propto G^{-2}\ ,
\end{equation}
\begin{equation}\label{1}
r\propto G^{-1}\ ,
\end{equation}
and
\begin{equation}\label{2}
v\propto G\ .
\end{equation}
Eq.(2) implies that as long as $G$ is constant then $T, r$, and $
v$ are constant too. However, there is a possibility that $G$
might have been changing appreciably over cosmological time. In
this case if one knows the way how $G$ varies the variation of the
distance $r$ can be calculated. Thus the variation of $G$ will
mimic the  tidal effects which people now attribute these changes
to. If $G$ changes with time Newton's gravitational law still
holds. However, the equivalence principle of general theory of
relativity is broken. The variation of $G$ may not be real and it
is due to an existence of dark matter coexisting with normal
matter. Its effect is to make the gravitational coupling (Newton's
constant) appear to be increasing. The effect of a little normal
matter and increasing gravity in a universe is equivalent to that
of more matter and normal gravity universe. We may dictate that
Newton's law of gravitation (and Kepler's law) to be applied to an
evolving local system, like the planetary system, viz.
Earth-Moon-Sun system.

In our present study we rely on a general form for the variation
of $G$ with cosmic time (Arbab, 2003). In this scenario a
gravitating body interacts with an effective gravitational
constant $G_{\rm eff.}$ which differs from a bare Newton's
constant we used to know. We have, in particular, an increasing
$G=G_{\rm eff.}$ at the present epoch, viz.
\begin{equation}\label{4}
G_{\rm eff.}=G_0\left(\frac{t}{t_0}\right)^n
\end{equation}
where $n>0$ is some constant to be determined from experiment. In
this sense gravity couples with $G_{\rm eff.}$ rather than with
bare $G_0$ because of cosmic expansion. The effect of this
constant is to replace in all formulae the normal(bare) Newton's
constant with this effective constant. The Earth couples with the
rest of the universe with this value. This coupling follows from
the idea of Mach that the inertia of an object is influenced by
the rest of matter in the universe. In an evolving universe this
effective constant induces a cosmological effect on over planetary
system while the bare constant ($G_0$) stays invariant. This why
we observe some cosmic effects exhibited in tidal effects, or
effects drawn from perturbation by other objects in the nearbye
solar system. In this context one has a calculable variation  in
the strength of gravity due to cosmic expansion. This variation
can't be measured directly. We present here a new approach of
detecting its variation with cosmic time in the way it has
affected planetary system dynamics. An increasing gravitational
constant may mimic an increasing mass of a gravitating body. Or
alternatively, it mimics a dark matter nearbye the gravitating
body that makes the orbiting object to fall towards it. A universe
with increasing gravitational constant may look indistinguishable
from the one with dark matter. Hence, if gravity increases for
some reason the idea of dark matter need not be attractive.
Milgrom modified Newton's law to account for the flattening of the
rotation curve. In our present case the modification does not
change the form of Newton's law.

Here $n$ determines the properties of the cosmological model
proposed. If one assumes that the length of the year remains
constant, then the length of the day $(D)$ should scale as
\begin{equation}\label{1}
D\propto G_{\rm eff.}^{2}\ ,
\end{equation}
Hence, one has
\begin{equation}\label{1}
T_0D_0=TD\ ,
\end{equation}
where the subscript `0' on the quantity denotes its present value.
Our model shows that the day was six hours when the earth was
formed. The angular velocity of the earth about the sun is
($\Omega=\frac{2\pi}{T}$)
\begin{equation}\label{1}
\Omega\propto G_{\rm eff.}^2
\end{equation}
This implies that the Earth is accelerating at a rate of
\begin{equation}\label{1}
\frac{\dot\Omega}{\Omega}=2\frac{\dot G_{\rm eff.}}{G_{\rm eff.}}
\end{equation}
and at the same time the earth-sun distance decreases at  a rate
of
\begin{equation}\label{5}
\frac{\dot r}{r}=-\frac{\dot G_{\rm eff.}}{G_{\rm eff.}}\
\end{equation}
If we know how $G$ varies, one can calculate this variation. In
our cosmological model, we know the general  variation of $G$
depends on a parameter $n$ that determines the whole cosmology. If
$n$ is know then the whole cosmological parameters are known. Our
model [1] could not determine $n$ exactly. It places a weaker
limit on the value of $n$. However, Wells [2] had found from a
palaeontological study the number of days in the year. To
reproduce his result we require the age of the universe to be
$t_0\sim 11$ billion years and $n=1.3$ so that
\begin{equation}\label{}
 G_{\rm eff.}=G_0\left(\frac{t}{t_0}\right)^{1.3}
\end{equation}
Hence, eqs.(4)-(7)  become, respectively
\begin{equation}\label{}
T=T_0\left(\frac{t_0}{t}\right)^{2.6}\ ,
\end{equation}
\begin{equation}\label{}
r=r_0\left(\frac{t_0}{t}\right)^{1.3}\ ,
\end{equation}
\begin{equation}\label{}
v=v_0\left(\frac{t}{t_0}\right)^{1.3}\ ,
\end{equation}
and
\begin{equation}\label{}
D=D_0\left(\frac{t}{t_0}\right)^{2.6}\ .
\end{equation}
We remark here that there is an astrophysical system (Binary
Pulsars) in which the decay of orbit is very prominent and
attributed to the emission of gravitational waves. Can we assume
here that there is a similar effect? Alternatively, may one
suggest that the decay of orbit in the former system is due to
cosmic expansion in the manner we have identified above? This is
quite plausible if the apparent acceleration of planetary system
is the direct cause.

It is worth to mention that Wells could not go far beyond the
Precambrian (600 million years back). Our model gives a formula
that determines the number of days in a year and the length of day
at any time in the past. These data obtained from this formula are
in full agreement with those obtained by different methods (see
Arbab 2004 and references therein). The correctness of the formula
entitle us to say that our initial proposition that the expansion
of the universe affects our solar system is correct. If that is
true, one can calculate the present acceleration of the earth (and
other planets) and the rate at which their orbit decreases. For
the earth orbital motion one finds that the present acceleration
amounts to
\begin{equation}\label{1}
\dot\Omega_0=\left(\frac{2.6}{t_0}\right)\ \Omega_0 \ \ , \ \
\dot\Omega_0=3.05\ \rm arcsec/cy^2
\end{equation}
At the same time the earth-sun distance decreases at a a rate of
\begin{equation}\label{5}
\dot r_0=-\left(\frac{1.3}{t_0}\right)r_0\ \ , \ \ \dot r_0=-\rm
17.7 m/year\ .
\end{equation}
One can also write the above equations in terms of Hubble
constant, as
\begin{equation}\label{2}
 \frac{\dot\Omega}{\Omega}=2.36\ H\ ,\qquad \frac{\dot
 r}{r}=-1.18\ H\ , \qquad \frac{\dot v}{v}=1.18 \ H\ ,\qquad \frac{\dot
 D}{D}=2.36\ H\ ,
\end{equation}
since
\begin{equation}\label{1}
\frac{\dot G_{\rm eff.}}{G_{\rm eff.}}=1.18 \ H \ .
\end{equation}
We see that the gravitational force increases as
\begin{equation}\label{}
F=\left(\frac{G_{\rm eff.}}{G_0}\right)^3F_0\ ,
\end{equation}
and upon using eq.(7), it becomes
\begin{equation}\label{}
F=\left(\frac{t}{t_0}\right)^4F_0\ ,
\end{equation}
We notice that the Newton's and Kepler's laws of gravitation do
still work well, even in an expanding universe, with only a minor
generalization that takes care of time evolution. The increase of
the gravitational forces is such that to counteract the present
universal expansion (acceleration) so that the universe remains in
equilibrium (flat). The gravitational force between our Earth and
the Sun  4.5 billion years ago has been $12\%$ less than now.

An increasing gravity would mean that in the past the gravity was
weak. This might probably provide a comfortable life of gigantic
animal, like dinosaurs, to roam freely on earth's surface. As
gravity increases their wights would become heavier and finally
may not support its growing weight. Thus it might not have been
appropriate for them to survive and later they vanish when they
are overweight. This scenario may provide a rather convenient
mechanism on how dinosaurs extinct.

We thus see that all earth parameters vary as due to universe
expansion. We have found the present Hubble constant to be
$H_0=10^{-10}\rm y^{-1}$ so that $\left(\frac{\dot G_{\rm
eff.}}{G_{\rm eff.}}\right)_0=1.18\times 10^{-10}\rm y^{-1}$. This
analysis imposes a new limit on $G_{\rm eff.}$ and $H$ which can
be tested with observational data. There is only few models that
deal with increasing $G$. However, models in which $G$ decreases
with time lead to serious difficulties, when confronted with
observations. Our model predicts a universal acceleration of all
gravitating bodies. For instance, we found that Mercury
accelerates at a rate of $12.6 \rm \ arcsec/cy^2$, Venus at a rate
of $4.95\ \rm arcsec/cy^2$, Earth at a rate of $3.05 \rm \
arcsec/cy^2$ and Mars at a rate of $1.6\ \rm arcsec/cy^2$. We
remark that the formulae pertaining to the planets motion are in
good agreement with observation. We should also await the
emergence of new data to test their applicability to these
systems. We see from eq.(19) that the day ($D$) lengthens by 1.95
msec/cy. According to scale expanding cosmos (SEC) the planets
spins down.  If all of these data are found to be in accordance
with observation, then our hypothesis that the cause of the
present acceleration is due to gravity increase would be
inventible!
\section{Concluding Remarks}
We have shown in this paper that the present cosmic acceleration
induces its effect on our Earth-Moon-Sun system. This is apparent
in the magnitude of the variation of the length of day, year,
distance, angular velocity, etc which are all proportional to the
Hubble parameter. The cosmological effects show up  in different
forms some of them are understood as due to tidal effects. We
anticipate that the future observations will bring many puzzles
and surprises  with it.
\newpage
\begin{table}[t]
\begin{center}
\caption{Data obtained  from  fossil corals and radiometric time}
\vspace{0.5cm}
\begin{tabular}{|r|r|r|r|r|r|r|r|r|r|}
\hline
Time$^a$  & 65 & 136 & 180 & 230 & 280 & 345 & 405 & 500 & 600\\
\hline
 days/year & 371.0 & 377.0 & 381.0 & 385.0 & 390.0 & 396.0 & 402.0 & 412. 0 & 424.0\\
\hline
\end{tabular}
\end{center}
\end{table}
\begin{table}[h]
\begin{center}
\caption{Data obtained from our empirical formula in eqs.(14) and
(17)} \vspace{0.5cm}
\begin{tabular}{|r|r|r|r|r|r|r|r|r|r|}
\hline
Time\footnote{in million years before present}   & 65 & 136 &  180  & 230 & 280  & 345  & 405 & 500 & 600 \\
\hline
 days/year  & 370.9 & 377.2 & 381.2 & 385.9 & 390.6 & 396.8 & 402.6 & 412.2  & 422.6\\
\hline
day (hr) & 23.6 & 23.2 & 23.0 & 22.7 & 22.4 & 22.1 & 21.7 & 21.3 & 20.7\\
\hline
Time$^a$   & 715 & 850 & 900  & 1200  & 2000 & 2500 & 3000 & 3560 & 4500\\
\hline
days/year  & 435.0 & 450.2 & 456  & 493.2  & 615.4 & 714.0 & 835.9 & 1009.5 & 1434\\
\hline
 day (hr) & 20.1  & 19.5 & 19.2  & 17.7  & 14.2& 12.3 & 10.5 & 8.7 & 6.1\\
\hline
\end{tabular}
\end{center}
\end{table}
$^a$ Time in million years before present\\
\section{References}
Arbab, A.I., \textbf{Class. Quantum Gravit}.23, 23 (2003)\\
Arbab, A.I., \textbf{Acta Geod.Geophys.Hung}.39, 27 (2004)\\
Arbab, A.I., \textbf{Acta Geod.Geophys.Hung.}40, 33 (2005 )\\
Kolesnik, Y. B.,  \textbf{Proceedings of the IAU} (2000)\\
Kolesnik, Y. B. and Masreliez C. J., \textbf{Astronomical Journal}.128, 878 (2004)\\
Wells, J.W.,  \textbf{Nature}.197, 948 (1963)\\
Masreliez, C.J,  \textbf{Apeiron}, V.11, 1 (2004)\\
\end{document}